\documentclass[conference]{IEEEtran}
\IEEEoverridecommandlockouts
\usepackage{color}
\usepackage{graphicx}  
\usepackage{psfrag}    
\usepackage{algorithm}
\usepackage{algorithmic}
\usepackage{amsfonts,amssymb}
\usepackage{amsthm}
\usepackage{amsmath}
\usepackage{amstext}
\usepackage{bm}
\usepackage{subfigure}
\usepackage{hyperref}
\usepackage{CJK}
\usepackage{url}
\usepackage{diagbox}
\usepackage[square, comma, sort&compress, numbers]{natbib}

\begin{document}

\title{A Cooperative Content Dissemination Framework for Fog-Based Internet of Vehicles}

\author{Weihua Wu, Peng Wang, Yuan Zhang, Weijia Han, He Yi and Tony Q. S. Quek, \emph{Fellow, IEEE}
\thanks{*This work was supported in part by the NSF China under Grant 61801365, 61701365 and 61971327, in part by the National Natural Science Foundation of Shaanxi Province under Grant 2019JQ-152, in part by Postdoctoral Foundation in Shaanxi Province of China, and the Fundamental Research Funds for the Central Universities.}
\thanks{W. Wu, and W. Han are with the School of Physics and Information Technology, Shaanxi Normal University, Xi'an 710119, China. (Email:whwu@snnu.edu.cn).}
\thanks{P. Wang and Yuan Zhang is with State Key Laboratory of ISN, School of Telecommunications Engineering, Xidian University, No.2 South Taibai Road, Xi'an, 710071, Shaanxi, China.}
\thanks{He Yi is with Experimental Training Base of National University of Defense Technology, Shaanxi Civil-ilitary Integration Key Laboratory of Intelligent Collaborative Network, doc.he@163.com.}
\thanks{T. Q. S. Quek is with the Information Systems Technology and Design Pillar, Singapore University of Technology and Design, Singapore 487372 (e-mail: tonyquek@sutd.edu.sg).}

}
\maketitle


\begin{abstract}
As the fog-based internet of vehicles (IoV) is equipped with rich perception, computing, communication and storage resources, it provides a new solution for the bulk data processing. However, the impact caused by the mobility of vehicles brings a challenge to the content scheduling and resource allocation of content dissemination service. In this paper, we propose a time-varying resource relationship graph to model the intertwined impact of the perception, computation, communication and storage resources across multiple snapshots on the content dissemination process of IoV. Based on this graph model, the content dissemination process is modeled as a mathematical optimization problem, where the quality of service of both delay tolerant and delay sensitive services are considered. Owing to its NP-completeness, the optimization problem is decomposed into a joint link and subchannel scheduling subproblem and as well a joint power and flow control subproblem. Then, a cascaded low complexity scheduling algorithm is proposed for the joint link and subchannel scheduling subproblem. Moreover, a robust resource management algorithm is developed for the power and flow control subproblem, where the channel uncertainties in future snapshots are fully considered in the algorithm. Finally, we conduct simulations to show that the effectiveness of the proposed approaches outperforms other state-of-art approaches.

\vspace{5pt}
\textbf{\emph{Key Terms}}: Internet of vehicles, content dissemination, robust resource optimization, uncertain channel.

\end{abstract}


%

\section{Introduction}

\newtheorem {theorem}{\textbf{Theorem}}
\newtheorem {lemma}{\textbf{Lemma}}
\newtheorem {remark}{\textbf{Remark}}
\newtheorem {definition}{\textbf{Definition}}

Nowadays, vehicles are equipped with a large number of perception devices, such as accelerators, radars, cameras, and advanced data processing units. With these abilities, vehicles can provide multiple location-based services, such as real-time map building \cite{KitaeRAL}, traffic management \cite{HlaingGCCE}, crowdsensing \cite{YangACCESS}, and environmental monitoring \cite{WenzhongTVT,MasahikoTITS}, etc,. With the development of intelligent transportation system (ITS), a large number of perceptual data is continuously generated \cite{WendiTVT,LuningTVT}. It is estimated that if 25\% of all vehicles are connected, 400 million GB of data will be transmitted every month \cite{LuningTVT}. In traditional ITS, tasks are often offloaded to cloud centers or edge servers via cellular networks, as shown in Fig. 1(a). As a result, it is challenging to tackle the dissemination problem for such a large amount of task content. On the one hand, the sheer amount of data transmission may exhaust the bandwidth resources of the cellular networks, which will lead to a large network delay. On the other hand, the base station (BS) provides wireless access service for the vehicles with LTE technology. It is costly to upload the massive data through BS.

\begin{figure}
\begin{center}
\includegraphics[width=3.0in,height=3.4in]{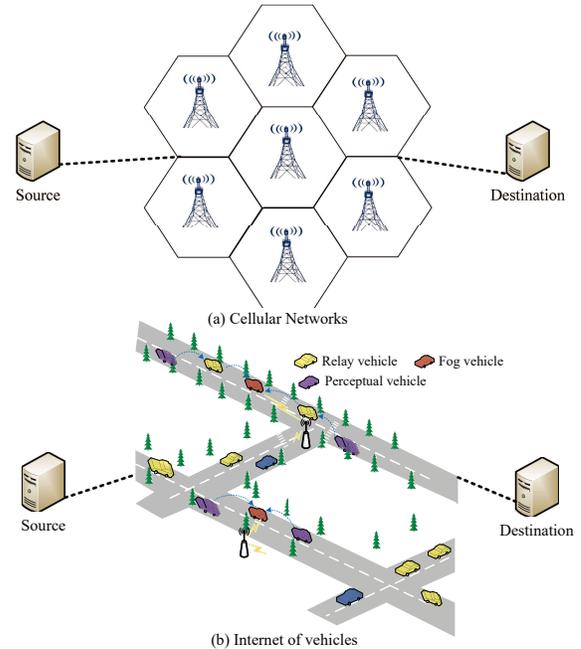}
\label{fig2}\caption{(a) Content dissemination based on cellular network architecture. (b) Content dissemination based on internet of vehicles.}
\end{center}
\end{figure}

Thanks to the vehicular fog computing technology, it is possible to process the whole life cycle of tasks at the internet of vehicles (IoV). As shown in Fig. 1(b), different vehicles will cooperate with each other to complete object perception and content dissemination.  Specifically, the perceptual vehicle will first sense the object, then the relay vehicle will store and forward the content, then the fog vehicle will compute and compress the content, and finally the computing results will be transmitted back to the requesters. Through the collaboration of the above components,  the IoV can be considered as a data processing platform between data source and destination \cite{MingZhaoITNSE}, avoiding many network infrastructure upgrades and bringing more bandwidth resources.

In general, fog computing platforms can be deployed on slower-moving commercial vehicles, such as buses and taxis, to handle service requests generated by neighboring vehicles or passengers \cite{ZhuCHAOIoT}. As a result, the tasks generated by vehicles can be handled by the fog nodes without having to be offloaded to the BS. In this way, the load of the BS and the delay of the task can be greatly reduced. However, it also adds a layer of complexity to the content dissemination process in IoV. Firstly, due to the high-speed movement of vehicles, the connection topology of vehicles changes dynamically. For high-volume tasks, it is difficult to complete their transmission in a single topology snapshot. Secondly, for the sake of achieving the content dissemination, we require to guarantee that the communication link is active not only at present but also in the next few slots. However, the wireless channel states in next few slots are unknown, which brings difficulties to the wireless resource management in IoV.  Thirdly, at the vehicle level, a vehicle can be within the communication range of multiple vehicles, but the efficient Vehicle-to-Vehicle (V2V) protocol only supports one-to-one transmission. Meanwhile, the wireless subchannel can only be allocated to one communication link in a time slot, leading to an NP-hard link scheduling and subchannel allocation problem. Finally, within the IoV, there also exist communications for security traffic such as collision warning, lane change request, etc. That means that while supporting delay-tolerant large volume services, the content dissemination scheme cannot affect the operation of security services in the IoV. This introduces a new design difficulty.

This paper proposes a cooperative content dissemination framework for IoV systems. We extend the conventional time-expanded graph to characterize the intertwined effects of the information perception, transmission, carry and computing resources on the content dissemination process of IoV. Based on the above extended graph model, the content dissemination process is formulated as a mathematical optimization problem, with the consideration of flow equilibrium constraint and the wireless resource constraint.
Because this optimization problem is NP-complete, it can be decomposed into a joint link and subchannel scheduling problem and as well a joint power and flow control problem. It should be pointed out that all possible communication paradigms, i.e., connected forwarding, carry-and-forward and direct forwarding, that can be implemented in vehicular networks are considered in the solving of the joint link and subchannel scheduling problem. Moreover, various communication requirements, such as the capacity constraint for volume content dissemination and the reliability constraint for emergency communication, are incorporated in the design of joint link and subchannel scheduling algorithm. The main contributions of this thesis can be summarized as follows:

\begin{itemize}
  \item  A time-varying resource relationship graph (TRRG) is proposed to characterize the time-varying coordination among perception, communication, storage and computation resources and their intertwined effects on the content dissemination process in IoV.
  \item We develop a cascaded joint link and subchannel scheduling algorithm. To be specific, the conflict relationship among different links is modeled as a conflict graph and the link scheduling problem is converted into finding maximal weighted independent sets from the conflict graph. Then, Hungarian algorithm is employed for the subchannel allocation among different links.
  \item A robust joint power allocation and flow control algorithm is developed for the resource management in IoV. Specifically, a learning-based robust transformation approach is developed to transform the chance-constrained power optimization problem into a convex deterministic form. With the given power allocation solution, the flow control is converted into a convex optimization problem that can be solved using the standard optimization tools.
\end{itemize}

The remainder of this paper is organized as follows. Section II reviews the related work and Section III presents the system model and the problem formulation. In Section IV, we discuss the link and subchannel scheduling problem. The power and flow control problem is presented in Section V. Simulation results are presented in Section VI. Finally, Section VII concludes the paper.

\section{Related Work}


There has been much current research on the content dissemination for IoV networks.
The work in \cite{GuiyangLuoChina} modeled the communication topology of IoV as a graph, and the content dissemination process was formulated as a maximum weight independent set (MWIS) problem. The content dissemination scheme in \cite{ChenchenTMC} allowed the vehicles to cooperate with their neighbors to complete the dissemination of popular content. A delay-sensitive routing algorithm was developed in \cite{XiaojieTVT} for the content dissemination in real-time traffic management system. However, these works consider that the BS disseminates the content to vehicles within the coverage range, and does not make full use of the communication resources between vehicles. The works in \cite{ChaoIoT,JindouXieChinacom,SunGangIoT} considered the computation-rich vehicles as cluster heads and all vehicles within its coverage transmit data to this cluster head, but they didn't consider the content dissemination based on carry-and-forward transmission model in IoV. In contrast, the works in \cite{MalandrinoTMCl,AlamTVTl,OptimalTMC,MalandrinoTVT} considered a variety of transfer paradigms, including  carry-and-forward, and then modeled the content dissemination problem as a flow scheduling problem on a dynamic network topology graph. However, they have not explored the wireless resource management in the vehicle networks.


Only appropriate wireless resource management can effectively support the content dissemination services. Thus, an optimal time allocation solution algorithm was proposed in \cite{ZipengTMC} for the data collection in content dissemination service.
In order to improve vehicle quality of service (QoS) and network utilisation, \cite{liuINFOCOM} proposed an effective network selection and traffic assignment approach.
The work in \cite{M.I.AshrafGlobecom} devised a novel proximity and load-aware resource allocation approach for V2V communication to minimize the total network cost.
An energy sensing-based resource allocation algorithm was developed in \cite{P.WangTWC} for ensuring the spectrum sharing between Vehicle-to-Infrastructure (V2I) and V2V users.
Nevertheless, it should be noted that all of above works propose the resource management approaches based on the assumption that the perfect instantaneous channel state information (CSI) is available at transmitters. In fact, it is difficult to complete the transmission process for the content dissemination service in a single topology snapshot. This means that the resource management algorithm must reserve resources to support the content dissemination in the next few timeslots. The strategies in \cite{L.LiangTCOM}\cite{C.GuoTVT} can make resource management decisions based on the distribution law of network states. However, it is difficult for them to adapt to the changes of the distribution of network states in the future.

\section{System Model}

\begin{figure}
\begin{center}
\includegraphics[scale = 0.8]{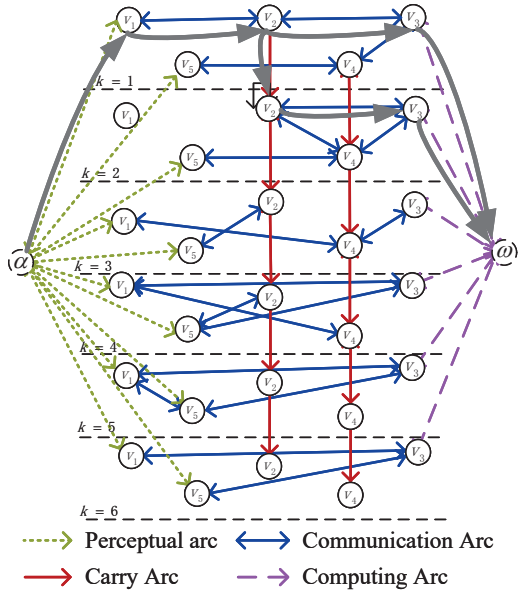}
\label{fig2}\caption{Time-varying resource relationship graph for the IoV, where $V_{1}$ and $V_{5}$ are perceptual vehicles, $V_{2}$ and $V_{4}$ are relay vehicles, $V_{3}$ is fog vehicle, $\alpha$ and $\beta$ are the virtual vertexes. The solid black lines indicate possible content dissemination paths.}
\end{center}
\end{figure}

\subsection{Network Model}
The IoV completes the content dissemination service according to the scheduling results provided by the traffic management server \cite{XiaojieTVT}.
First, the perceptual vehicles capture the raw images of the observation targets, i.e. traffic accidents, road damages, traffic jams, etc., when they are within the observable range of the on-board sensors.
Then, the raw images should be transmitted to the fog vehicle for multi-source data fusion and image understanding. Because of vehicle movements, the links between different vehicles can only be established when they move into each other's coverage area. This means that the images are delivered to the fog vehicle through the relay vehicle. Moreover, in order to transmit more data or to take advantage of better opportunities in the future, the relay vehicles can also store data first and then forward it upon moving to the better locations. The image is compressed after the fog vehicle has received the data. It's not difficult to understand that the compressed image will be very small so that it can be relayed to the requester via BS. In the whole process of content dissemination, the large data transmission occurs at the vehicle level, and  the BS is only responsible for the transmission of compressed results. Therefore, this architecture can significantly reduce the load of BS. The perceptual, relay and fog vehicles are encouraged by the incentive mechanism to join the content dissemination service. Some vehicles on the road do not join the content dissemination service due to their selfishness or limited resources. We denote them as audience vehicles (AVs).

We denote the set of perceptual vehicles as $\mathcal{V}_{u}$, the set of fog vehicles  as $\mathcal{V}_{c}$ and the set of relay vehicles as $\mathcal{V}_{r}$.
We denote the set of all vehicles as $\mathcal{V}=\mathcal{V}_{u}\bigcup \mathcal{V}_{c}\bigcup \mathcal{V}_{r}$. To better differentiate vehicles, we use $\mathcal{V}_{e}=\mathcal{V}_{c}\bigcup \mathcal{V}_{r}$ to represent the set of vehicles that have no data to upload and $\mathcal{V}_{n}=\mathcal{V}_{u}\bigcup \mathcal{V}_{r}$ to denote the set of vehicles without installing the computing platform.
There are $\mathcal{S}$ tasks that are perceived by the perceptual vehicles and handled by the fog vehicles.
We consider all possible communication paradigms in vehicular networks.

\emph{Connected forward}, resulting from the V2V links between different vehicles, represents the typical way in which vehicles communicate with each other in vehicular networks.

\emph{Carry-and-forward}, the relay vehicle stores and carries the received data and waits until it is within the communication range of the destination vehicle, and then forwards the data.

\emph{Direct forward}, which refers to the direct communication between BS and vehicles, represents the typical way for vehicles to communicate with the infrastructure.

As we know, for high-volume data transmission, it is difficult to accomplish it in a single topology snapshot. However, different from the traditional wireless networks, the mobility trajectory of vehicles can be accurately predicted by many learning methods \cite{TangYujieTVT,WangXiuminTVT,ZhuXiaoyuTITS}. Thus, the network topology in the future can be constructed according to the predicted trajectory.  The time-extended graph (TEG) \cite{Exploring} is an efficient method to model the connection relationship in the network across multiple snapshots. In this paper, we extend the traditional TEG to model the intertwined impact of the object perception, transmission, carry and computating resources on the content dissemination process of IoV as a time-varying resource relationship graph (TRRG) $\mathcal{G}_{K}$. Fig. 2 illustrates an example TRRG of the vehicular networks with five vehicles, where $V_{1}$ and $V_{5}$ are  perceptual vehicles, $V_{2}$ and $V_{4}$ are relay vehicles, $V_{3}$ is fog vehicle. $\alpha$ and $\beta$ are the virtual vertexes, which represent the perceptual data source and the computing unit on the fog vehicle. TRRG is a directed graph composed multiple layers. Each layer corresponds to a contact event. At the beginning of each event, the link between any two vehicles is established. On the contrary, one or ones links are removed at the end of this event. Thus, a topology snapshot is extracted from each event.  The time interval of one contact event in the network is called a frame. Within each frame, the network is considered as static. Thus, TRRG can be used to approximate a continuously evolving vehicular network where the network topology is static in each frame and changes only at frame transitions.

We assume that the TRRG $\mathcal{G}_{K}$ consists of $K$ layers.
We represent the i-th vehicle participating in the network at layer (frame) $k$ as $v_{i}^{k}$.
There are two types of vertexes in TRRG: ordinary vertexes and virtual vertexes, which represent the temporal copies of vehicles and the virtual sink or source, respectively.
There are four types of arcs in $\mathcal{G}_{K}$ to characterize the different resources in vehicular network.

\emph{Communication Arc:} If a link between the non-fog vehicle $v_{i}^{k}$ and the non-perceptual $v_{j}^{k}$ is active during that frame, then a directed arc $(v_{i}^{k},v_{j}^{k})$ exists from vertex $v_{i}^{k}\in \mathcal{V}_{n}^{k}$ to vertex $v_{j}^{k}\in \mathcal{V}_{e}^{k}$. The set of arcs with destination of vertex $v_{j}^{k}$ is represented as $\mathcal{L}_{v,j}^{k}$. Within frame $k$, the set of communication arcs is $\mathcal{L}_{v}^{k}=\bigcup\limits_{v_{j}^{k}\in \mathcal{V}_{e}^{k}} \mathcal{L}_{v,j}^{k}$.

\emph{Computing Arc:} At any frame $k$, there is always an arc $(v_{i}^{k},\omega)$ from fog vertex $v_{i}^{k}\in \mathcal{V}_{c}^{k}$ to virtual node $\omega$. The set of computing arcs is represented as $\mathcal{L}_{c}$.

\emph{Perception Arc:} At any frame $k$, there is always an arc $(\alpha,v_{s}^{k})$ \footnote{$v_{s}^{k}$ refers to the s-th perceptual vehicle that perceives task s.} from virtual node $\alpha$ to perception vertex $v_{s}^{k}\in \mathcal{V}_{u}^{k}$, the upload rate is $\mu(\alpha,v_{s}^{k},s)$ . The set of perception arcs is represented as $\mathcal{L}_{\mu}$.

\emph{Carry Arc:} A directed arc $(v_{i}^{k},v_{i}^{k+1})$ is also drawn from any vertex $v_{i}^{k}\in \mathcal{V}_{r}^{k}$ to vertex $v_{i}^{k+1}\in \mathcal{V}_{r}^{k+1}$.

We assume that the V2V links are supported by model-1 of NR sidelink. To improve the spectrum efficiency, the subchannels are reused by V2V links and AVs. The assignment of spectrum resource is denoted as the indicator variable $a_{(i,j)}^{m,k}$.  Specifically, $a_{(i,j)}^{m,k}=1$ when the V2V link between vehicle $i$ and $j$  reuses the spectrum of the $m$-th AV, and $a_{(i,j)}^{m,k}=0$ otherwise. Let $p_{(i,j)}^{k}$ and $p_{m}^{k}$ represent the transmit powers of link $(i,j)$ and the $m$-th AV. The channel power gain of link $(i,j)$ and the $m$-th AV at frame $k$ is denoted as $g_{(i,j)}^{k}$ and $g_{m}^{k}$, respectively. Then, the SINR of link $(i,j)$ at frame $k$ is
\begin{eqnarray}
\gamma_{(i,j)}^{k}=\frac{p_{(i,j)}^{k}g_{(i,j)}^{k}}{\sum_{m\in \mathcal{M}} a_{(i,j)}^{m,k}p_{m}^{k}g_{m}^{(i,j),k}+\sigma^{2}},  \label{1}
\end{eqnarray}
where $g_{m}^{(i,j),k}$ is the corsstalk channel gain from $m$-th AV to link $(i,j)$ and $\sigma^{2}$ is the power of the additive white Gaussian noise. Since the
channel gain $g_{(i,j)}^{k}$ and $g_{m}^{(i,j),k}$ change with time as well. Thus, we focus on the average rate of link  $(i,j)$ as
\begin{eqnarray}
\bar{c}_{i,j}^{k}(\gamma_{(i,j)}^{k})\!=\!\mathbb{E}\!\!\left[\!W\log_{2}\!\left(\!\!1\!+\!\frac{p_{(i,j)}^{k}g_{(i,j)}^{k}}{\sum_{m\in \mathcal{M}} a_{(i,j)}^{m,k}p_{m}^{k}g_{m}^{(i,j),k}\!\!+\!\sigma^{2}}\!\!\right)\!\!\right], \label{2}
\end{eqnarray}
where $W$ is the subchannel bandwidth. Similarly, the SINR of AV $m$ at frame $k$ is
\begin{eqnarray}
\gamma_{m}^{k}=\frac{p_{m}^{k} g_{m}^{k}}{\sum_{(i,j)\in \mathcal{L}_{v}^{k}} a_{(i,j)}^{m,k}p_{(i,j)}^{k}g_{(i,j)}^{m,k}+\sigma^{2}},  \label{3}
\end{eqnarray}
where $g_{(i,j)}^{m,k}$ is the corsstalk channel gain from link $(i,j)$ to the receiver of $m$-th AV.

After the fog vehicle compressing the data, the BS needs to transmit the computing results to the requester. In this process, the BS works a full duplex relay.
Specifically, the transmission rate from the fog vehicle to the BS can be expressed as
\begin{eqnarray}
R_{I}^{T_{s}}=W \log_{2}\left\{1+\frac{P_{i,o}^{T_{s}}G_{i,o}^{T_{s}}}{\sigma^{2}}\right\}, \forall v_{i}^{T_{s}} \in  \mathcal{V}_{c}^{T_{s}},s\in \mathcal{S},   \label{4}
\end{eqnarray}
where $T_{s}$ denotes the last frame of the transmission of task $s$, $P_{i,o}^{T_{s}}$ and $G_{i,o}^{T_{s}}$ are the transmit power and channel gain from the fog vehicle to the BS at frame $T_{s}$, respectively.  Similarly, the transmission rate from BS to the requesters can be expressed as
\begin{eqnarray}
R_{II}^{T_{s}}=W \log_{2}\left\{1+\frac{P_{o,i}^{T_{s}}G_{o,i}^{T_{s}}}{\sigma^{2}}  \right\}, \forall v_{i}^{T_{s}} \in  \mathcal{V}_{u}^{T_{s}},s\in \mathcal{S},  \label{5}
\end{eqnarray}
where $P_{o,i}^{T_{s}}$ and $G_{o,i}^{T_{s}}$ are the transmit power and channel gain from BS to requestor at frame $T_{s}$, respectively.

\subsection{Problem Formulation}

The basic idea of this work is to complete the task processing at the vehicle level as far as possible, so as to reduce the load of BS. To achieve this goal, we consider a network utility function for the TRRG, which is defined as the difference between the throughput of vehicles and the total energy consumption of BS
\begin{eqnarray}
\max \frac{1}{K} \sum_{k=1}^{K}\sum_{v_{s}^{k}\in V_{u}^{k}} \!\!U(\mu(\alpha,v_{s}^{k},s))-\!w_{p}\!\!\!\!\!\sum_{\iota \in \{I,II\} }\!\!\!\!(P_{\iota,o}^{T_{s}}+P_{o,\iota}^{T_{s}}),  \label{6}
\end{eqnarray}
where $w_{p}$ is the weight parameter between different objectives. The throughput of vehicles is represented as
\begin{eqnarray}
U(\mu(\alpha,v_{s}^{k},s))=\log(\mu(\alpha,v_{s}^{k},s)+e),  \label{7}
\end{eqnarray}
where $e$ is the base of natural logarithm. Based on this throughput function, the task flows can be scheduled to balance the congestion across multiple frames. Hereinafter, we introduce the following constraints for the optimization problem on TRRG.

\subsubsection{Nonnegative Flow Constraints} To ensure that all data flows in the vehicle network are nonnegative, we construct the following constraints. The upload flow from virtual source node to the perceptual vehicles should satisfy
\begin{eqnarray}
\mu(\alpha,v_{s}^{k},s)\geq 0, \forall s\in \mathcal{S}, k\in \mathcal{K}, v_{s}^{k}\in \mathcal{V}_{u}^{k}.  \label{8}
\end{eqnarray}
We define the flow of task $s\in \mathcal{S}$ on communication arc $(v_{i}^{k},v_{j}^{k})$ as $x(v_{i}^{k},v_{j}^{k},s)$, it should satisfy
\begin{eqnarray}
 x(v_{i}^{k},v_{j}^{k},s)\geq 0,\forall s\in \mathcal{S}, k\in \mathcal{K}, v_{i}^{k}\in \mathcal{V}_{n}^{k}, v_{j}^{k}\in \mathcal{V}_{e}^{k}.  \label{9}
\end{eqnarray}
The computing flow for task $s\in \mathcal{S}$ on computing arc $(v_{i}^{k},\omega)$ is presented as $d(v_{i}^{k},\omega,s)$. We make it greater than zero, i.e.,
\begin{eqnarray}
d(v_{i}^{k},\omega,s)\geq 0,\forall s\in \mathcal{S}, k\in \mathcal{K}, v_{i}^{k}\in \mathcal{V}_{c}^{k}. \label{10}
\end{eqnarray}
Finally, we denote the flow of $s\in \mathcal{S}$ on carry arc $(v_{i}^{k},v_{i}^{k+1})$ as $x(v_{i}^{k},v_{i}^{k+1},s)$ and it should satisfy
\begin{eqnarray}
x(v_{i}^{k},v_{i}^{k+1},s)\geq 0,\forall s\in \mathcal{S}, k\in \mathcal{K}, v_{i}^{k}\in \mathcal{V}_{r}^{k}. \label{11}
\end{eqnarray}

\subsubsection{Flow Balance Constraints} For any vertex on TRRG, the amount of input flows should be equal to the amount of output flows. This principle maps to different expressions for different vertices. For the perceptual vehicle $v_{s}^{k}\in \mathcal{V}_{u}^{k}$ at frame $k$, this maps to
\begin{eqnarray}
\mu(\alpha,v_{s}^{k},s)=\sum_{v_{i}^{k}\in \mathcal{V}_{e}^{k}}x(v_{s}^{k},v_{i}^{k},s), \forall s\in \mathcal{S}, k\in \mathcal{K}, v_{s}^{k}\in \mathcal{V}_{u}^{k}. \label{12}
\end{eqnarray}
For the relay vehicle $v_{i}^{k}\in \mathcal{V}_{r}^{k}$ at frame $k$, the flow balance constraint maps to
\begin{eqnarray}
&&\!\!\!\!\!\!\!\!\!\!\!\!\!\underbrace{\sum_{v_{j}^{k}\in \mathcal{V}_{n}^{k} }\sum_{s\in \mathcal{S}} x(v_{j}^{k},v_{i}^{k},s)}_{\textrm{From noncomputing node}}+\underbrace{\sum_{s\in \mathcal{S}} x(v_{i}^{k-1},v_{i}^{k},s)}_{\textrm{From carry arc}}= \label{13}\\
&&\!\!\!\!\!\!\!\!\!\!\!\!\!\!\!\!\!\!\!\!\!\underbrace{\sum_{v_{j}^{k}\in \mathcal{V}_{e}^{k} }\!\sum_{s\in \mathcal{S}} \!x(v_{i}^{k}\!,\!v_{j}^{k}\!,\!s)}_{\textrm{To nonuploading node}}\!+\!\underbrace{\sum_{s\in \mathcal{S}}\! x(v_{i}^{k}\!,\!v_{i}^{k+1}\!,\!s)}_{\textrm{To carry arc}}, \forall k\in \mathcal{K}, v_{i}^{k}\in \mathcal{V}_{r}^{k}. \nonumber
\end{eqnarray}

\subsubsection{Link Scheduling Constraints} 
There exist conflicts in the scheduling of the same resource, on account of the restriction of vehicle platform attitude. For example, because of using a single antenna, the vehicles can only communicate with only one vehicle at one frame, even if there are multiple vehicles in its communication range \cite{ChinaComLuo}. For modeling this kind of conflict, we introduce a set of boolean variables $\delta(v_{i}^{k},v_{j}^{k})\in \{0,1\}$, whose value is 1 if link $(v_{i}^{k},v_{j}^{k})$ is active at $k$ frame and 0 otherwise. Then the conflicts of communication resource can be formulated as
\begin{eqnarray}
\sum_{v_{j}^{k}\in \mathcal{V}_{e}^{k}}\delta(v_{i}^{k},v_{j}^{k})\leq 1,   \forall k\in \mathcal{K}, v_{i}^{k}\in \mathcal{V}_{n}^{k}, \label{14}
\end{eqnarray}
and
\begin{eqnarray}
\sum_{v_{j}^{k}\in \mathcal{V}_{n}^{k}}\delta(v_{j}^{k},v_{i}^{k})\leq 1,   \forall k\in \mathcal{K}, v_{i}^{k}\in \mathcal{V}_{e}^{k}, \label{15}
\end{eqnarray}
which indicate that vehicle $v_{i}^{k}$  can transmit data to at most one vehicle and  receive data from at most one vehicle at the same time, respectively. These two constraints indicate that the arc starting from the data source vehicle is a directed arc. Moreover, there also exist conflicts in the scheduling of communication resource of the relay vehicle, i.e., receiving and transmitting cannot be carried out at the same time, which can be expressed as
\begin{eqnarray}
\sum_{v_{j}^{k}\in \mathcal{V}_{n}^{k}}\delta(v_{j}^{k},v_{i}^{k})+ \sum_{v_{j}^{k}\in \mathcal{V}_{e}^{k}}\delta(v_{i}^{k},v_{j}^{k})\leq 1, \forall k\in \mathcal{K},  v_{i}^{k} \in \mathcal{V}_{r}. \label{16}
\end{eqnarray}

\subsubsection{Capacity Constraints} The flows carried on the communication arc cannot exceed its capacity, which can be expressed as
\begin{eqnarray}
\sum_{s\in \mathcal{S}} x(v_{i}^{k},v_{j}^{k},s)\leq &&\!\!\!\!\!\!\!\!\!\delta(v_{j}^{k},v_{i}^{k})\bar{c}_{i,j}^{k}(\gamma_{(i,j)}^{k}), \nonumber\\
 &&\forall k\in \mathcal{K}, v_{i}^{k} \in \mathcal{V}_{n}^{k}, v_{j}^{k}\in \mathcal{V}_{e}^{k}. \label{17}
\end{eqnarray}
The capacity constraint of carry flow is
\begin{eqnarray}
\sum_{s\in \mathcal{S}}x(v_{i}^{k},v_{i}^{k+1},s)\leq c_{ar}(v_{i}^{k}), \forall v_{i} \in \mathcal{V}_{r}, k\in \mathcal{K}, \label{18}
\end{eqnarray}
where $c_{ar}(v_{i}^{k})$ is the maximum cache capacity of vehicle $v_{i}^{k}$. For the fog vehicle $v_{i}^{k} \in \mathcal{V}_{c}^{k}$ at frame $k$, we can obtain
\begin{eqnarray}
\underbrace{\sum_{v_{j}^{k}\in \mathcal{V}_{n}^{k} }\sum_{s\in\mathcal{S}}x(v_{j}^{k},v_{i}^{k},s)}_{\textrm{From noncomputing node}}\leq d(v_{i}^{k},\omega,s),\forall k\in \mathcal{K}, v_{i}^{k} \in \mathcal{V}_{c}^{k}, \label{19}
\end{eqnarray}
where $d(v_{i}^{k},\omega,s)$  is the maximum computing capacity of the computing vehicle.

\subsubsection{Delay Constraints}  For task $s\in \mathcal{S}$, it must be transmitted in $T_{s}$ frames, i.e.,
\begin{eqnarray}
&&\!\!\!\!\!\!\!\!\!x(v_{i}^{k},v_{j}^{k},s)=0, \forall s\in \mathcal{S}, k\geq T_{s}, v_{i}^{k}\in \mathcal{V}_{n}^{k},v_{j}^{k}\in \mathcal{V}_{e}^{k},  \label{20}\\
&&\!\!\!\!\!\!\!\!\!x(v_{i}^{k},v_{i}^{k+1},s)=0, \forall s\in \mathcal{S}, k\geq T_{s}, v_{i}^{k}\in \mathcal{V}_{r}^{k}. \label{21}
\end{eqnarray}

\subsubsection{Computation Results Transmission}  The amount of data received by fog vehicles is $\sum_{k=1}^{T_{s}}\mu(\alpha,v_{s}^{k},s)$. The fog vehicle will call its image compression module to compress the received data. We assume that the fog vehicle can provide lossless compression with ratio as $\eta$.  Then the compression result needs to be transmitted to the requester by full duplex BS. In order to avoid outage \cite{DangTVT}, the transmission process needs to meet the following constraint
\begin{eqnarray}
\Pr\{R_{\iota}^{T_{s}}\geq \theta_{s}, \iota=I,II\}\geq 1-\epsilon, \forall s \in \mathcal{S}, \label{22}
\end{eqnarray}
where $1-\epsilon$ is the maximum tolerable outage probability and $\theta_{s}=\eta\sum_{k=1}^{T_{s}}\mu(\alpha,v_{s}^{k},s)$.

\subsubsection{Wireless Resource Constraints} 
The system needs to allocate enough wireless resources to the vehicles to effectively support the flows in TRRG. Therefore, it is necessary to determine the constraints on wireless resources. Firstly, the transmit powers of vehicles and BS should not exceed their maximum values, i.e.,
\begin{eqnarray}
&&p_{(i,j)}^{k}\leq P_{max}^{v},  \forall k\in \mathcal{K}, v_{i}^{k} \in \mathcal{V}_{n}^{k}, v_{j}^{k}\in \mathcal{V}_{e}^{k}, \label{23}\\
&&p_{m}^{k}\leq P_{m}^{max}, \forall k\in \mathcal{K}, m\in \mathcal{M},\label{24}\\
&&P_{i,o}^{T_{s}}\leq P_{i,o}^{max}, \forall v_{i}^{Ts} \in \mathcal{V}_{c}^{Ts},  \label{a1}\\
&&P_{o,i}^{T_{s}}\leq P_{o,i}^{max}, \forall v_{i}^{Ts} \in \mathcal{V}_{u}^{Ts}.  \label{a2}
\end{eqnarray}

It should be noted that the wireless channel states in future frames are unknown, which bring difficulties for the wireless resource management in IoV. To overcome this difficulty, we assume that the system can model the wireless channel states for a period of time in the future according to the vehicle trajectory and the known vehicle environment. Obviously, the modeled wireless channel states are not necessarily accurate. The content dissemination services that this paper focuses on are generally delay tolerant. However, it is very tricky that the traffic, e.g, vehicle platooning, advanced driving, remote driving, etc., in AVs often requires the communication to have very high reliability \cite{GarciaTCST}. In order to meet the high reliability requirements under uncertain channels, we give the following Quality of Service (QoS) constraints in probabilistic form for AVs,
\begin{eqnarray}
\Pr\{\gamma_{m}^{k} \geq \gamma_{v}^{th}\}\geq 1-\epsilon, \forall k \in \mathcal{K},   m \in \mathcal{M}. \label{25}
\end{eqnarray}
Besides that, the following subchannel allocation constraints
\begin{eqnarray}
   && a_{(i,j)}^{m,k}\in \{0,1\}, \forall k \in \mathcal{K}, m \in \mathcal{M}, (i,j)\in \mathcal{L}_{v}^{k}, \label{26} \\
   && \sum_{(i,j)\in \mathcal{L}_{v}^{k}}a_{(i,j)}^{m,k} \leq 1, \forall k\in \mathcal{K}, m \in \mathcal{M},  \label{27}  \\
   && \sum_{m \in \mathcal{M} }a_{(i,j)}^{m,k} \leq 1, \forall k\in \mathcal{K}, (i,j) \in \mathcal{L}_{v}^{k}, \label{28}
\end{eqnarray}
must be satisfied to indicate that the spectrum of one AV can only be shared with a single V2V link and one V2V link is only allowed to access the spectrum of a single AV.

Above all, considering all of the constraints and objective function, the problem of cooperative content dissemination can be written as
\begin{eqnarray}
\!\!\!\!\!\!\!\!\!\textrm{\textbf{P1}:}\!\! \max_{\bm\delta,\mathbf{a},\mathbf{x},\mathbf{p},\mathbf{P}} &&\!\!\!\!\!\!\!\!\!\!\!\! \sum_{k=1}^{K}\!\sum_{v_{s}^{k}\in V_{u}^{k}} \!\!\!U(\mu(\alpha,v_{s}^{k},s))\!-\!\!w_{p}\!\!\!\!\!\!\sum_{\iota \in \{I,II\}}\!\!\!(P_{\iota,o}^{T_{s}}\!+\!P_{o,\iota}^{T_{s}})  \label{29}\\
\textrm{s.t.} && \!\!\!\! (\ref{8})-(\ref{28}). \nonumber
\end{eqnarray}
In \textbf{P1}, $\bm\delta=[\delta(v_{i}^{k},v_{j}^{k})]_{(i,j,k)}$ and $\mathbf{a}=[a_{(i,j)}^{m,k}]_{(i,j,m,k)}$ are integer variables, $\mathbf{x}=[x(v_{j}^{k},v_{i}^{k},s)]_{(i,j,k,s)}$, $\mathbf{p}=[p_{(i,j)}^{k},p_m^k]_{(i,j,m,k)}$ and $\mathbf{P}=[P_{\iota,o}^{T_{s}},P_{o,\iota}^{T_{s}}]_{(\iota)}$ are continuous variables. Moreover, the constraint in Eq. (\ref{17}) is non-linear. Therefore, \textbf{P1} is a mixed integer non-linear programming (MINLP) problem, which is NP-hard in general \cite{RZLACCESS}.

\subsection{Problem Analysis and Decomposition}


From the graphical model, the power control variables $\mathbf{p}$, $\mathbf{P}$ and the flow variable $\mathbf{x}$ jointly control the volume of flow. In comparison, the subchannel allocation variable $\mathbf{a}$ and link scheduling variable $\bm\delta$ determine whether a communication arc is active or not. Moreover, both $\mathbf{a}$ and $\bm\delta$ are boolean variables. When their values are given, the remaining optimization problem about $\mathbf{p}$, $\mathbf{P}$ and $\mathbf{x}$ is continuous optimization problem. Motivated by the above discovery, \textbf{P1} in TRRG can be decomposed into two subproblems to decrease the complexity.

\begin{enumerate}
  \item \textbf{Link and subchannel scheduling problem (LSP)}: Solve $\mathbf{a}$ and $\bm\delta$ with given channel state information.
  \item \textbf{Resource allocation problem}: Solve $\mathbf{p}$, $\mathbf{P}$ and $\mathbf{x}$ with given link and subchannel scheduling solution.
\end{enumerate}

With the decomposition of \textbf{P1}, joint link and subchannel scheduling algorithm and joint power and flow control algorithm are developed in the following sections, respectively.

\section{Link and Subchannel Scheduling Problem }

The purpose of link scheduling is to obtain the set of active links. Therefore, before the channel assignment, we must determine the active links between vehicles. To achieve the joint link scheduling and subchannel allocation solution, we propose a cascaded  scheme. More specifically, the link scheduling of V2V is first conducted to determine the activation link, where the link scheduling constraints (\ref{14})-(\ref{16}) are satisfied. Based on the link schedule results, the allocation of subchannel is subsequently performed.
Our proposed cascaded scheme not only satisfies the constraints in (\ref{14})-(\ref{16}) and (\ref{26})-(\ref{28}), but also reduces the computational complexity.

\subsection{Link Scheduling}

The objective of link scheduling is to maximize the capacity of V2V. However, before the link scheduling, the optimal powers cannot be allocated, thus the capacity of V2V links cannot be calculated. To overcome this difficulty, the link scheduling is controlled to maximize the overall throughput of V2V links based on their Channel-to-Noise Ratios (CNRs) i.e., $g_{(i,j)}^{k}/\sigma^{2}$.

\begin{figure}
\begin{center}
\includegraphics[width=3.0in,height=1.8in]{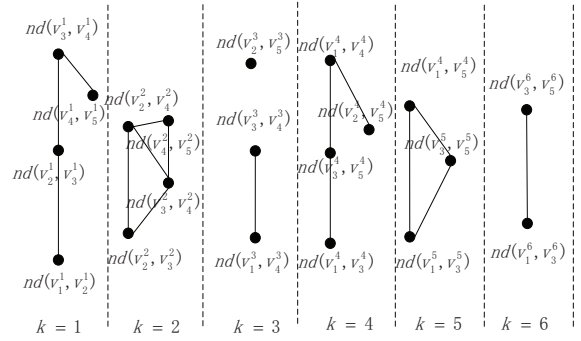}
\label{fig2}\caption{Conflict graph for the communication resources.}
\end{center}
\end{figure}


From the view of graph theory,  constraints (\ref{14})-(\ref{16}) describe the conflict relation of different communication arcs.
For achieving a conflict-free scheduling in TRRG, a conflict graph, designated by CG, is proposed to model the conflict relationship among different communication resources.
For example, Fig. 3 illustrates the conflict graph of the communication resources as shown in Fig. 2.
Each node in the conflict graph represents a possible resource scheduling solution which corresponds a communication arc in TRRG. For example, node $nd(v_{3}^{1},v_{4}^{1})$ in Fig. 3 represents the scheduling of the communication resource $(v_{3}^{1},v_{4}^{1})$ at the first frame in Fig. 2. The edges in the conflict graph represent the conflict relationship between the communication resources. In other words, if the two communication resources conflict with each other, there exists an edge between the two nodes corresponding to them in CG. Similar to the TRRG, the conflict graph is a layered graph.
Furthermore, since edges only connect nodes within the same layer, the layers of the conflict graph are independent of each other.
The resource scheduling contained in an independent set of the conflict graph are conflict-free, because there is no edge between the nodes of an independent set in CG. Hence, by sequentially finding independent sets for each layer of the conflict graph, the conflict-free scheduling of the communication arcs can be obtained. The maximum weighted independent set may be any one of all the maximal independent sets, which is related to the weight of each node. Therefore, in order to obtain the maximum weighted independent set, first all the maximal independent sets must be obtained, and then the maximum weighted independent set can be obtained by calculating the weighted sum of each maximal independent set. In this subsection, we obtain the maximal independent sets of the CG by using the Bron-Kerbosch algorithm \cite{bron1973algorithm} in a recursive manner. By calculating the weight, we can find the maximum weight independent set from all the maximum independent sets.

\subsection{Hungarian Algorithm Strategies}

After the link scheduling in above subsection, the optimal set of communication links is activated. Subsequently, appropriate subchannels need to be allocated to these links to ensure that they can effectively support the content flows. It is not difficult to see that the channel allocation between V2V and AV can be regarded as a bipartite matching problem in graph theory. The Hungarian algorithm is an effective bipartite matching optimization algorithm for the subchannel allocation problem \cite{C.GuoTVT}.  However, before the resource allocation, the optimal powers cannot be allocated thus the capacity of V2V links cannot be calculated. Given that our objective is to maximize the throughput of TRRG, we assume that the V2V links have priorities to access channels. Let $\varphi_{(i,j)}^{m}=g_{m}^{(i,j),k}/\sigma^{2}$ denote the interference link CNR for the $m$-th AV over the link $(i,j)$. Therefore, channels are assigned to V2V according to the Hungarian algorithm to minimize the total CNR of the interference links from AVs to V2Vs and
thus to minimize the co-channel interference.

\section{Power and Flow Control}

Through the proposed cascaded scheduling algorithm, not only the activation link set is obtained but also the V2V and AV who share the same channel form a reusing pair. Then, the resource allocation problem can be carried out under the given TRRG structure.

It is not difficult to understand that any vertex can only be connected to one communication arc at any frame. Under this solution, the flow balance constraints in (\ref{12}) (\ref{13}) and the communication capacity constraint in (\ref{17}) (\ref{19}) can be rewritten as 
\begin{eqnarray}
&&\!\!\!\!\!\!\!\!\!\!\!\!\!\!\mu(\alpha,v_{s}^{k})=\delta^{*}(v_{s}^{k},v_{i}^{k}) x(v_{s}^{k},v_{i}^{k},s), \forall s\in \mathcal{S}, k\in \mathcal{K}, \label{30}  \\
&&\!\!\!\!\!\!\!\!\!\!\!\!\!\!\sum_{s\in \mathcal{S}}\delta^{*}(v_{j}^{k},\!v_{i}^{k}) x(v_{j}^{k},v_{i}^{k},s)+\sum_{s\in \mathcal{S}} x(v_{i}^{k-1},v_{i}^{k},s)=  \label{31} \\
&&\!\!\!\!\!\!\!\!\!\!\!\!\!\!\sum_{s\in \mathcal{S}}\delta^{*}(v_{i}^{k},\!v_{j}^{k})\! x(v_{i}^{k}\!,\!v_{j}^{k}\!,\!s)\!+\!\sum_{s\in \mathcal{S}}\! x(v_{i}^{k}\!,\!v_{i}^{k+1}\!,\!s), \forall s,k, v_{i}^{k}\in \mathcal{V}_{r}^{k}, \nonumber\\
&&\!\!\!\!\!\!\!\!\!\!\!\!\!\!\sum_{s\in\mathcal{S}}\delta^{*}(v_{j}^{k},v_{i}^{k})x(v_{j}^{k},v_{i}^{k},s)\leq d(v_{i}^{k},\omega),\forall s,k, v_{i}^{k} \in \mathcal{V}_{c}^{k}, \label{32}   \\
&&\!\!\!\!\!\!\!\!\!\!\!\!\!\!\sum_{s\in \mathcal{S}} \!x(v_{i}^{k}\!,\!v_{j}^{k}\!,\!s)\!\leq\! \delta^{*}\!(v_{j}^{k}\!,\!v_{i}^{k})\bar{c}_{i,j}^{k}(\gamma_{i,j}^{k}), \!\forall s,k,\!v_{i}^{k}\! \in \!\mathcal{V}_{n}^{k}\!,\!v_{j}^{k}\!\in\!\mathcal{V}_{e}^{k}, \label{33}
\end{eqnarray}
where $\delta^{*}$ is the link scheduling solution obtained in above section.

Under the subchannel allocation solution, the wireless resource constraint at each frame $k\in \mathcal{K}$ can be reformulated as
\begin{eqnarray}
&&\!\!\!\!\!\!\!\!\!\!\!\!\!\!\Pr\left\{\frac{p_{m}^{k}g_{m}^{k}}{a_{(i,j)}^{m,k,*}p_{(i,j)}^{k}g_{(i,j)}^{m,k}+\sigma^{2}}\!\geq\!\gamma_{v}^{th}\right\}\!\geq\! 1\!-\!\epsilon,  \forall k ,  m, \label{34}
\end{eqnarray}
where $a^{*}$ represents the subchannel allocation solution.

After link scheduling and subchannel allocation, constraints (\ref{14})-(\ref{16}) and (\ref{26})-(\ref{28}) can be eliminated.  Then by replacing (\ref{12}) (\ref{13}) (\ref{17}) (\ref{19}) and (\ref{25}) with (\ref{30})-(\ref{33}) and (\ref{34}), respectively, the cooperative content dissemination problem can be reformulated as
\begin{eqnarray}
\!\!\!\!\!\!\!\!\textrm{\textbf{P2}:}  \max_{\mathbf{x},\mathbf{p},\mathbf{P}} &&\!\!\!\!\!\!\!\!\! \sum_{k=1}^{K}\!\sum_{v_{s}^{k}\in V_{u}^{k}} \!\!\!U(\mu(\alpha,v_{s}^{k}))\!-\!w_{p}\!\!\sum_{\iota \in \{I,II\}}(P_{\iota,o}^{T_{s}}\!+\!P_{o,\iota}^{T_{s}})  \label{35}\\
\textrm{s.t.} &&\!\!\!\!\!\! (\ref{8})-(\ref{11}),(\ref{18}),(\ref{20})-(\ref{24}),(\ref{30})-(\ref{34}).  \nonumber 
\end{eqnarray}
Although the integer variables have been eliminated, problem \textbf{P2} is still difficult to solve.  First, the chance constraints in (\ref{22}) and (\ref{34}) pose a great difficult on computing the optimal solution of \textbf{P2}. Second, the expectation in Eq. (\ref{2}) makes Eq. (\ref{17}) a statistical constraint. Third, the interference in Eq. (\ref{2}) makes \textbf{P2} a nonconvex optimization problem. All of these bring difficulties for solving \textbf{P2}. In following subsections, we will overcome these difficulties separately.

\subsection{Learning-based robust counterpart for (\ref{34})}

The chance constraint in (\ref{34}) is caused by the uncertain CSI of AVs. Rather than directly resolving the chance constraint, this subsection proposes a robust optimization approach to represent the uncertain CSI by high-probability-region (HPR) and enforces the inequality constraint in (\ref{34}) to hold for any uncertain CSI within it. To acquire the HPR of uncertain CSI, we should collect multiple independent and identically distributed (i.i.d) samples of the CSI to learn the uncertainty set.
In this subsection, we use a convex set to cover the uncertainties of the CSI. If the obtained power allocation solutions are feasible under the constructed uncertainty set, all of the constraints can be satisfied.

Inspired by this consideration, we reformulate constraint (\ref{34}) as
\begin{eqnarray}
 p_{m}^{k}g_{m}^{k}/\gamma_{v}^{th}-a_{(i,j)}^{m,k,*}p_{(i,j)}^{k}g_{(i,j)}^{m,k} \geq \sigma^{2}, \mathbf{g} \in \mathcal{G}, \label{36}
\end{eqnarray}
where $\mathbf{g}=\{g_{m}^{k}/\gamma_{v}^{th},a_{(i,j)}^{m,k,*}g_{(i,j)}^{m,k}\}$ and $\mathcal{G}$ is the HPR that needs to be learned. It is not difficult to understand that by choosing $\mathcal{G}$ to cover a $1-\epsilon$ content of $\mathbf{g}$, i.e., $\Pr\{\mathbf{g}\in \mathcal{G}\}\geq 1-\epsilon$, any resource allocation solution that satisfies (\ref{36}) must satisfy
\begin{eqnarray}
\Pr\{p_{m}^{k}g_{m}^{k}/\gamma_{v}^{th}-a_{(i,j)}^{m,k,*}p_{(i,j)}^{k}g_{(i,j)}^{m,k} \geq \sigma^{2}\}\geq 1-\epsilon. \label{37}
\end{eqnarray}

Then, motivated from the tractability of the resulting robust optimization \cite{gorissen2015practical}, we use the ellipsoid  set to model the uncertainties of the channel realizations. Thus, the HPRs of  $\mathcal{G}$ can be parameterized as
\begin{eqnarray}
\mathcal{G}=\{\mathbf{g}:\mathbf{g}=\bar{\mathbf{g}}+\mathbf{B}\mathbf{u},\mathbf{u}^{T}\mathbf{u}\leq1\}, \label{38}
\end{eqnarray}
where $\mathbf{B}\in\mathbb{R}^{2\times2}$ and $\mathbf{u}\in\mathbb{R}^{2}$. Here, $\bar{\mathbf{g}}$ and $\mathbf{B}$ are the parameters that should be learned from the sample data sets.

In this subsection, a statistical leaning method is proposed to obtain the parameters of HPR.  First, we need to collect continuous i.i.d sample data sets for $\mathbf{g}$ as $\mathcal{D}=\{\bm\xi_{1},\bm\xi_{2},\cdots,\bm\xi_{D}\}$, where $\bm\xi_{i}\in \mathbb{R}^{2}$. Understandably, the ellipsoid set $\mathcal{G}$ can be reparameterized as
\begin{eqnarray}
\mathcal{G}=\{\mathbf{g}: (\mathbf{g}-\bar{\mathbf{g}})^{T}\bm\Lambda^{-1}(\mathbf{g}-\bar{\mathbf{g}})\leq z_{e} \}, \label{39}
\end{eqnarray}
where $\bar{\mathbf{g}}$ is the center of $\mathcal{G}$, $z_{e}>0$ is the size of $\mathcal{G}$ and $\bm\Lambda\in\mathbb{R}^{2\times2}$ determines the relationships between different channel realizations. Without loss of generalization, the center of $\mathcal{G}$ can be chosen as the sample mean, i.e.
\begin{eqnarray}
\bar{\mathbf{g}}=\frac{1}{D} \sum_{l=1}^{D} \bm\xi_{l}, \label{40}
\end{eqnarray}

We chose $\bm\Lambda$ as the covariance matrix, which can be computed as
\begin{eqnarray}
\Lambda_{i,j}=\frac{1}{D}\sum_{l=1}^{D} (\bm\xi_{l}^{(i)}-\bar{\mathbf{g}}^{(i)}) (\bm\xi_{l}^{(j)}-\bar{\mathbf{g}}^{(j)}), \label{41}
\end{eqnarray}
where $i=1,2$ and $j=1,2$.

Then, we need to calibrate the uncertainty sets so that they satisfy the chance constraint $\Pr\{\mathbf{g}\in \mathcal{G}\}\geq 1-\epsilon$. For calibrating uncertainty set $\mathcal{G}$, we estimate the $1-\epsilon$ quantile of data samples in $\mathcal{D}$.  Let
\begin{eqnarray}
t(\bm\xi)=(\bm\xi-\bar{\mathbf{g}})^{T}\bm\Lambda^{-1}(\bm\xi-\bar{\mathbf{g}}) \label{42}
\end{eqnarray}
be the map from the random space $\mathbb{R}^{2}$ into $\mathbb{R}$. Based on the data samples in $\mathcal{D}$, we define the $(1-\epsilon)$-quantile $q_{1-\epsilon}$ of the underlying distribution of $t(\bm\xi)$ from
\begin{eqnarray}
\Pr\{t(\bm\xi)\leq q_{1-\epsilon}\}=1-\epsilon. \label{43}
\end{eqnarray}
By computing the function values of $t(\bm\xi)$ on each sample of $\mathcal{D}$, we can obtain the observations $t(\bm\xi_{(1)}),\cdots,t(\bm\xi_{(D)})$. Then, the $k^{*}_{c}=\lceil(1-\epsilon)D\rceil$-th value of the ranked observations $t_{(1)}\leq\cdots\leq t_{(D)}$ in ascending order can be considered as the upper bound of $(1-\epsilon)$-quantile of $t(\bm\xi)$. As a result, the size of uncertainty set $\mathcal{G}$ can be set as
\begin{eqnarray}
z_{e}=t(\bm\xi_{(k^{*}_{c})}). \label{44}
\end{eqnarray}
Based on these results, matrixe $\mathbf{B}$ can be computed as
\begin{eqnarray*}
\mathbf{B}=\sqrt{z_{e}}\bm\Delta ,
\end{eqnarray*}
where $\bm\Delta$ are the Cholesky decompositions of $\bm\Lambda$, i.e. $\bm\Lambda=\bm\Delta \bm\Delta^{T}$. Then, we can summarize the whole procedure for the learning of HPRs $\mathcal{G}$ in \textbf{Algorithm 1}.

\begin{algorithm}[h]\label{Am1}
\caption{Statistical Learning Approach for Uncertainty Sets}
\begin{algorithmic}[0]
\STATE \textbf{Input:} The sample channel gain sets $\mathcal{D}=\{\bm\xi_{1},\bm\xi_{2},\cdots,\bm\xi_{D}\}$;\\
\quad \textbf{Shape Learning:} Set shape parameters $\bar{\mathbf{g}}$ as Eq. (\ref{40}),  \\
\quad  $\bm\Lambda$ as Eq. (\ref{41}); \\
\quad \textbf{Size Calibration:} Set size parameter  $z_{e}$ as $t(\bm\xi_{(k^{*}_{c})})$;   \\
\quad Compute $\mathbf{B}=\sqrt{z_{e}}\bm\Delta$ through Cholesky \\
\quad decomposition $\bm\Sigma=\bm\Delta\bm\Delta^{T}$;\\
\STATE \textbf{Output:} $\bar{\mathbf{g}}$ and $\mathbf{B}$.
\end{algorithmic}
\end{algorithm}

Based on the ellipsoid uncertainty set, the constraint in (\ref{34}) holds if and only if $\mathbf{P}_{3}^{*}\geq \sigma^{2}$, where $\mathbf{P}_{3}^{*}$ is the optimum of the following optimization problem
\begin{eqnarray}
\textbf{P3}:\min_{\mathbf{g}} && \mathbf{p}^{T}\mathbf{g}  \label{45} \\
\textrm{s.t.} && \mathbf{g}=\bar{\mathbf{g}}+\mathbf{B} \mathbf{u},\mathbf{u}^{T}\mathbf{u}\leq 1. \label{46}
\end{eqnarray}
We refer to \textbf{P3} as the subproblem which must be solved. Since $\inf_{\parallel\mathbf{u}\parallel\leq 1}  \mathbf{p}^{T}(\bar{\mathbf{g}}+\mathbf{B} \mathbf{u})=\mathbf{p}^{T}\bar{\mathbf{g}}-\|\mathbf{p}^{T} \mathbf{B} \|$, where the derivation is based on the Schwartz inequality, the V2V QoS constraint in (\ref{34}) can be replaced by
\begin{eqnarray}
\mathbf{p}^{T}\bar{\mathbf{g}}- \|\mathbf{p}^{T}\mathbf{B}\|  \geq \sigma^{2}, \label{47}
\end{eqnarray}
which is a second-order cone. Thus, it is effectively compatible with the convex optimization tools.

\subsection{Joint Chance Constraint in Eq. (\ref{22})}

By using the Bonfreeoni's \cite{2013Dis} inequality, the joint chance constraint in (\ref{22}) can be converted into the following equation
\begin{eqnarray}
&&\Pr\left(\left\{R_{I}^{T_{s}}\geq \theta_{s}\right\}\bigcap \left\{R_{II}^{T_{s}}\geq\theta_{s}\right\}\right)\geq 1-\epsilon  \label{48} \\
&&\quad   \Longleftrightarrow \Pr\left(\left\{R_{I}^{T_{s}}\leq \theta_{s}\right\}\bigcup \left\{R_{II}^{T_{s}}\leq\theta_{s}\right\}\right)\leq\epsilon.   \nonumber
\end{eqnarray}
Furthermore, Bonferroni's inequality is equivalent to
\begin{eqnarray}
&&\Pr\left(\left\{R_{I}^{T_{s}}\leq \theta_{s}\right\}\bigcup \left\{R_{II}^{T_{s}}\leq\theta_{s}\right\}\right) \label{49}\\
&&\quad \leq \Pr\left(R_{I}^{T_{s}}\leq \theta_{s}\right)+\Pr\left(R_{II}^{T_{s}}\leq\theta_{s}\right).  \nonumber
\end{eqnarray}
For any vector of tolerable outage probability $\mathcal{E}=\{\bm\epsilon\in \mathbb{R}_{+}^{2}: \epsilon_{I}+\epsilon_{II}\leq \epsilon\}$, the following chance constraint
\begin{eqnarray}
\Pr\{R_{\iota}^{T_{s}}\geq \theta_{s}\}\geq 1-\epsilon_{\iota},  \iota=I,II \label{50}
\end{eqnarray}
represents a conservative approximation for the chance constraint in (\ref{22}). The problem of finding the best $\bm\epsilon$ is nonconvex and believed to intractable \cite{2013Dis}. As a result, in most applications of Bonferroni's inequality the tolerable outage probability $\epsilon$ is equally divided among the multiple chance constraints in (\ref{22}) by setting $\epsilon_{\iota}=\epsilon/2$ for $ \iota=I,II$. Note that the two transmission processes of the full duplex BS are independent of each other. By using a statistical method similar to \textbf{Algorithm 1}, the constraints can be equivalently transformed into
\begin{eqnarray}
&&W \log_{2}\left\{1+\frac{P_{i,o}^{T_{s}}G_{i,o}^{k_{d}^{*},T_{s}}}{\sigma^{2}}\right\}\geq \theta_{s}   \label{51}
\end{eqnarray}
and
\begin{eqnarray}
&&W \log_{2}\left\{1+\frac{P_{o,i}^{T_{s}}G_{o,i}^{k_{d}^{*},T_{s}}}{\sigma^{2}}  \right\}\geq \theta_{s}, \label{52}
\end{eqnarray}
where $k_{d}^{*}$ represent the $(1-\epsilon_{\iota})$ quantile of the channel gain. Thus, after the above steps, the joint chance constraint in (\ref{22}) is transformed into several solvable convex constraints.

\subsection{Expectation Constraint in Eq. (\ref{33})}

For the expectation of V2V transmission capacity,  we get the approximation based on lemma 1 in \cite{2014Power} as
\begin{eqnarray}
\bar{c}_{i,j}^{k}(\gamma_{(i,j)}^{k})&&\!\!\!\!\!\!\!\!\!\!\!=\!\mathbb{E}\!\!\left[\!W\log_{2}\!\left(\!\!1\!+\!\frac{p_{(i,j)}^{k}g_{(i,j)}^{k}}{p_{m}^{k}g_{m}^{(i,j),k}\!\!+\!\sigma^{2}}\!\!\right)\!\!\right] \label{53}\\
&&\!\!\!\!\!\!\!\!\!\!\!\approx \!W\log_{2}\!\left(\!\!1\!+\!\frac{p_{(i,j)}^{k}\mathbb{E}[g_{(i,j)}^{k}]}{p_{m}^{k}\mathbb{E}[g_{m}^{(i,j),k}]\!\!+\!\sigma^{2}}\!\!\right) \nonumber\\
&&\!\!\!\!\!\!\!\!\!\!\!=c_{i,j}^{k}(\bar{\gamma}_{(i,j)}^{k}), \nonumber
\end{eqnarray}
where both the expectation $\mathbb{E}[g_{(i,j)}^{k}]$ and $\mathbb{E}[g_{m}^{(i,j),k}]$ can be obtained by using the same learning method in Eq. (\ref{40}). Then, the expectation constraint in Eq. (\ref{17}) can be approximated as
\begin{eqnarray}
\!\!\!\!\!\sum_{s\in \mathcal{S}} x(v_{i}^{k},v_{j}^{k},s)\!\leq &&\!\!\!\!\!\!\!\!\!\!c_{i,j}^{k}(\bar{\gamma}_{(i,j)}^{k}),\forall k\!\in\! \mathcal{K}, v_{i}^{k}\!\in\! \mathcal{V}_{n}^{k}, v_{j}^{k}\!\in \!\mathcal{V}_{e}^{k}. \label{54}
\end{eqnarray}

\subsection{Decomposition of Resource Allocation}

Based on the learning and approximation methods in above subsections, we can transform optimization \textbf{P2} into
\begin{eqnarray}
\textrm{\textbf{P4}:}\max_{\mathbf{x},\mathbf{p},\mathbf{P}} &&\!\!\!\!\!\!\!\!\!\!\!\!\sum_{k=1}^{K}\!\sum_{v_{s}^{k}\in V_{u}^{k}} \!\!\!U(\mu(\alpha,v_{s}^{k}))\!-\!w_{p}\!\!\!\!\!\sum_{\iota \in \{I,II\}}\!\!\!(P_{\iota,o}^{T_{s}}\!+\!P_{o,\iota}^{T_{s}})  \label{55}\\
\textrm{s.t.} &&\!\!\!\!\!\! (\ref{8})-(\ref{11}),(\ref{18}),(\ref{20})-(\ref{21}),(\ref{47}),  \nonumber \\
&&\!\!\!\!\!\!(\ref{23})-(\ref{a2}), (\ref{30})-(\ref{32}),(\ref{51})-(\ref{52}),(\ref{54}). \nonumber
\end{eqnarray}
In the above problem, the flow control variables and power allocation variables are still coupled, which brings difficulty to the analysis of the problem. To decouple them, we construct the following Lagrangian function
\begin{eqnarray*}
L(\mathbf{x},\mathbf{p},\mathbf{P},\lambda)\!=&&\!\!\!\!\!\!\!\!\!\!\!\max_{\mathbf{x},\mathbf{p},\mathbf{P},\lambda}
\frac{1}{K} \!\sum_{k=1}^{K}\!\!\sum_{v_{s}^{k}\in V_{u}^{k}} \!\!U(\mu(\alpha,v_{s}^{k}))\!  \\
&&\!\!\!\!\!\!\!\!\!\!\!\!-w_{p}\!\!\!\!\sum_{\iota \in \{I,II\}}\!\!\!\!(P_{\iota,o}^{T_{s}}\!+\!P_{o,\iota}^{T_{s}}) \\
&&\!\!\!\!\!\!\!\!\!\!\!\!-\!\!\!\!\!\sum_{k,v_{i}^{k},v_{j}^{k}}\!\!\!\lambda_{k,v_{i}^{k},v_{j}^{k}}(\sum_{s\in \mathcal{S}} x(v_{i}^{k},v_{j}^{k},s)-c_{i,j}^{k}(\bar{\gamma}_{(i,j)}^{k})) \nonumber \\
\textrm{s.t.} &&\!\!\!\!\!\! (\ref{8})-(\ref{11}),(\ref{18}),(\ref{20})-(\ref{21}),(\ref{47}),  \nonumber \\
&&\!\!\!\!\!\!(\ref{23})-(\ref{a2}), (\ref{30})-(\ref{32}),(\ref{51})-(\ref{52}), \nonumber
\end{eqnarray*}
where $\lambda_{k,v_{i}^{k},v_{j}^{k}}$ is the Lagrange multiplier associated with (\ref{54}).

In the above problems, $U(\mu(\alpha,v_{s}^{k},s))$ and $x(v_{i}^{k},v_{j}^{k},s)$ are only related to flow control variables $\mathbf{x}$, $c_{i,j}^{k}(\bar{\gamma}_{(i,j)}^{k})$ is related to power allocation variables $\mathbf{p}$, and $\mathbf{P}$ is related to the full duplex communication of BS.  According to this observation, we can get the following power allocation problem about $\mathbf{p}$
\begin{eqnarray}
\textrm{\textbf{P5}:}\max_{\mathbf{p}}&&\!\!\!\!\!\!\!\!\!\!\!\!\sum_{k,v_{i}^{k},v_{j}^{k}}\lambda_{k,v_{i}^{k},v_{j}^{k}}c_{i,j}^{k}(\bar{\gamma}_{(i,j)}^{k}) \label{56}\\
\textrm{s.t.} &&\!\!\!\!\!\! (\ref{23})-(\ref{24}),(\ref{47}).  \nonumber
\end{eqnarray}
Through the Hungarian subchannel allocation in Section III, the V2V and AV who share the same subchannel form a V2V-AV pair. Since the different V2V-AV pairs adopt orthogonal subchannels, the power allocation in \textbf{P5} can be carried out at each single V2V-AV pair. In this case, $\lambda_{k,v_{i}^{k},v_{j}^{k}}$ only works as a coefficient, which has no impact on the solution of the problem. Then, the power allocation of link $(i,j)$-AV $m$ pair at frame $k$ is formulated as
\begin{eqnarray}
\textrm{\textbf{P6}:}\max_{p_{(i,j)}^{k},p_{m}^{k}}&&\!\!\!\!\!\!\!\!\!W\log_{2}\!\left(\!\!1\!+\!\frac{p_{(i,j)}^{k}\bar{g}_{(i,j)}^{k}}{p_{m}^{k}\bar{g}_{m}^{(i,j),k}\!\!+\!\sigma^{2}}\!\!\right) \label{57}\\
\textrm{s.t.} &&\!\!\!\!\!\!\mathbf{p}^{T}\bar{\mathbf{g}}- \|\mathbf{p}^{T}\mathbf{B}\|  \geq \sigma^{2},  \nonumber \\
&&\!\!\!\!\!\!  p_{(i,j)}^{k}\leq P_{max}^{v}, \nonumber\\
&&\!\!\!\!\!\!  p_{m}^{k}\leq P_{m}^{max}. \nonumber
\end{eqnarray}
Although problem \textbf{P6} is a non-convex optimization, it is a common problem related to the spectral reuse in IoV. This problem has been discussed several times in our previous work \cite{WuTWC} and a bisection search algorithm has been designed to obtain its optimal solution.

\begin{algorithm}[h]\label{Am1}
\caption{Bisection Search for Solving \textbf{P6}}
\begin{algorithmic}[0]
\STATE Set termination threshold $0<\zeta<1$;
\STATE Set $p_{m,min}^{k}=0$ and $p_{m,max}^{k}=P_{m}^{max}$;
\WHILE {$p_{m}^{k}<P_{m}^{max}-\zeta$}
\STATE set $p_{m}^{k}=(p_{m,min}^{k}+p_{m,max}^{k})/2$; Solve
\begin{eqnarray}
\textrm{\textbf{P7}:}\max_{p_{(i,j)}^{k}}&&\!\!\!\!\!\!\!\!\!W\log_{2}\!\left(\!\!1\!+\!\frac{p_{(i,j)}^{k}\bar{g}_{(i,j)}^{k}}{p_{m}^{k}\bar{g}_{m}^{(i,j),k}\!\!+\!\sigma^{2}}\!\!\right) \label{58}\\
\textrm{s.t.} &&\!\!\!\!\!\!\mathbf{p}^{T}\bar{\mathbf{g}}- \|\mathbf{p}^{T}\mathbf{B}\|  \geq \sigma^{2},  \nonumber \\
&&\!\!\!\!\!\!  p_{(i,j)}^{k}\leq P_{max}^{v}. \nonumber
\end{eqnarray}
 to obtain $p_{(i,j)}^{k}$;
\IF{$p_{(i,j)}^{k}>P_{max}^{v}+\zeta$}
\STATE $p_{m,max}^{k}=p_{m}^{k}$
\ELSIF{$p_{(i,j)}^{k}<P_{max}^{v}-\zeta$}
\STATE $p_{m,min}^{k}=p_{m}^{k}$
\ELSIF{$P_{max}^{v}-\zeta<p_{(i,j)}^{k}<P_{max}^{v}+\zeta$}
\STATE break
\ENDIF
\ENDWHILE
\STATE Output the optimal transmit powers $p_{(i,j)}^{k,*}$ and $p_{m}^{k,*}$.
\end{algorithmic}
\end{algorithm}

According to the given power allocation solution, the link capacity between V2Vs can be determined. Then, under the given link allocation, the joint flow control and BS power allocation problem can be formulated as
\begin{eqnarray}
\!\!\!\!\!\!\textrm{\textbf{P8}:}\max_{\mathbf{x},\mathbf{P}} &&\!\!\!\!\!\!\!\!\!\!\!\!\frac{1}{K} \sum_{k=1}^{K}\!\!\sum_{v_{s}^{k}\in V_{u}^{k}}\!\! U(\mu(\alpha,v_{s}^{k}))\!-\!w_{p}\!\!\!\!\!\!\sum_{\iota \in \{I,II\}}\!\!\!\!(P_{\iota,o}^{T_{s}}\!+\!P_{o,\iota}^{T_{s}}) \label{59} \\
\textrm{s.t.}&&\!\!\!\!\!\! (\ref{8})-(\ref{11}),(\ref{18}),(\ref{20})-(\ref{21}),(\ref{a1})-(\ref{a2}),  \nonumber \\
&&\!\!\!\!\!\!(\ref{30})-(\ref{32}),(\ref{51})-(\ref{52}),(\ref{54}). \nonumber
\end{eqnarray}
Notice that all constraints in \textbf{P8} are convex sets. Even more, its objective is a convex function. Thus, \textbf{P8} is a convex optimization problem and it can be effectively solved by the widely used convex optimization tools.

\section{Simulation Results}
\begin{figure}
\begin{center}
\includegraphics[width=3.8in,height=1.8in]{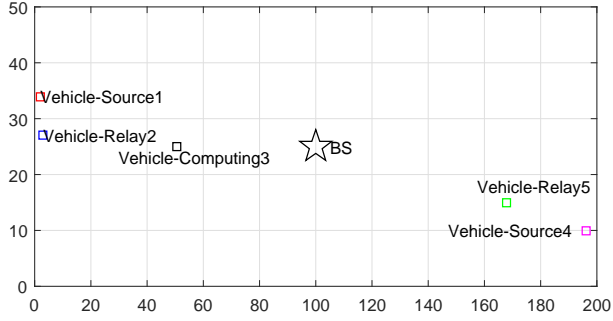}
\label{fig2}\caption{Vehicular network scenario.}
\end{center}
\end{figure}

In this section, we conduct the simulation to verify the performance of our proposed approaches. We consider a vehicular network scenario as shown in Fig. 4. Specifically, vehicles 1 and 4 are the perceptual vehicles, vehicles 2 and 5 are the relay vehicles and as well vehicle 3 is the fog vehicle. These vehicles are distributed in two lanes, and each lane is 25 meters wide. The BS is located in the middle of the lane. The initial coordinates of these components are given in Table I. Vehicles 1, 2 and 3 travel from left to right at speeds of 72, 108 and 18 kilometers per hour, respectively. Vehicles 4 and 5 travel from right to left at speeds of 140 and 72 kilometers per hour, respectively.  The other major simulation parameters and the channel model for IoV are described in Table II. In the simulation, we compare our proposed approach with four baseline approaches. The first one is the \emph{V2-Only} approach. In this approach, only the cache of vehicle v2 can be called by the content dissemination service. The second is the \emph{V5-Only} approach. Similar to \emph{V2-Only}, this approach only allows to call the cache of vehicle v5. The above two approaches use part of the storage resources in the IoV. In order to analyze the benefits of storage resources, we also run a \emph{Without-Carry} approach in the simulation, which cannot use the storage resources of any vehicle. In addition, we compare our proposed approach with the \emph{Non-Robust} approach as a baseline, where the power allocation problem is solved based on the average channel gain $\bar{\mathbf{g}}$.

\begin{table}[]
    \caption{Component Coordinates [m]}
    \centering
\begin{tabular}{|c|c|c|c|c|c|}
  \hline
  BS & v1 & v2 & v3 & v4 & v5 \\
  \hline
  (100,25) & (2,34) & (3,27) & (50.5,25) & (196,10) & (168,15) \\
  \hline
  \diagbox{}{} & AV1 & AV2 & AV3 & AV4 & AV5 \\
  \hline
  \diagbox{}{} & (20,45) & (120,50) & (190,35) & (100,5) & (0,25) \\
  \hline
\end{tabular}
\end{table}

\begin{table}[]
    \caption{Simulation Parameters}
    \centering
    \begin{tabular}{|c|c|}
        \hline
          \textbf{Parameter} & \textbf{Value} \\
        \hline
        RB bandwidth, $B$   & 10 MHz   \\
        \hline
        Noise spectrum density, $\sigma^{2}$   & -174 dBm/Hz  \\
        \hline
        Reliability for AVs, $\epsilon$   & $10^{-3}$  \\
        \hline
        Sample number for channel training, $N$   & 1000  \\
        \hline
        Pathloss model  & $128.1+37.6\log_{10}d[km]$  \\
        \hline
        Shadowing standard deviation  & 4 dB  \\
        \hline
        Fast fading  &  Rayleign fading  \\
        \hline
        Bisection search accuracy, $\zeta$   & $10^{-3}$  \\
        \hline
       Maximum power, $P_{m}^{max},P_{max}^{v},P_{i,o}^{max}$   & $30dBm$  \\
        \hline
       Compression with ratio, $\eta$   & $0.1$  \\
        \hline
    \end{tabular}
    \label{bs2}
\end{table}

\begin{figure}
\begin{center}
\includegraphics[width=3.6in,height=4in]{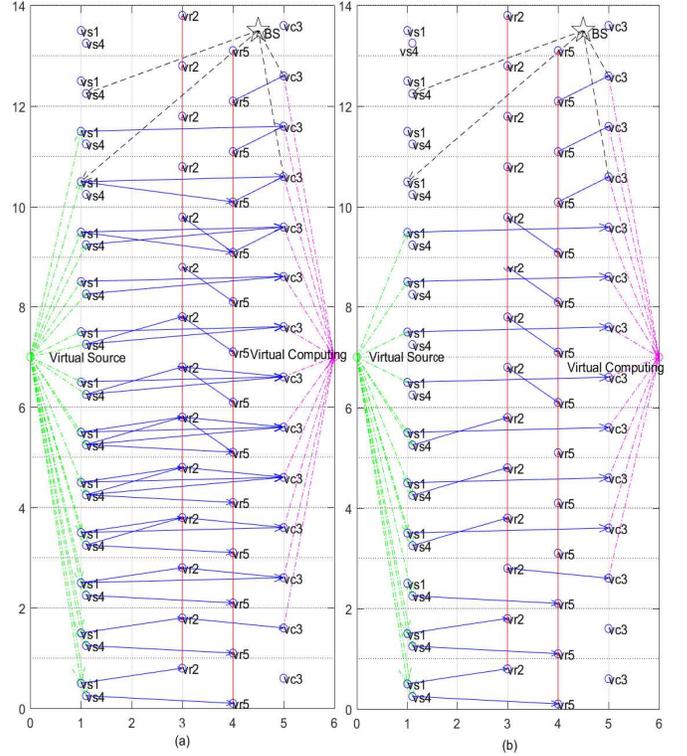}
\label{fig2}\caption{Conflict graph for the communication resources. (a) Before link scheduling. (b) After link scheduling}
\end{center}
\end{figure}

\begin{figure}
\begin{center}
\includegraphics[width=2.8in,height=2.2in]{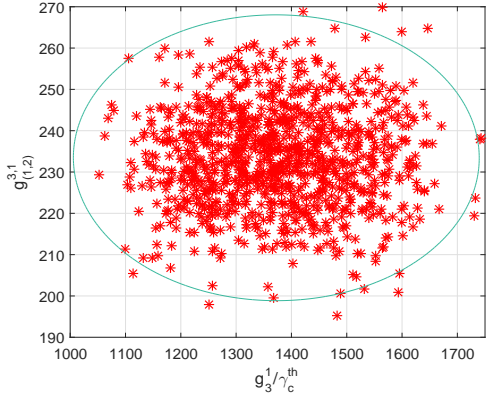}
\label{fig2}\caption{Construction of ellipsoid uncertain set.}
\end{center}
\end{figure}

Fig. 5 shows the  conflict relationship of the communication resources in IoV. Specifically, Fig.5 (a) shows the resource coupling relationship before link scheduling. It can be seen from the figure that there are multiple potential transmission paths\footnote{The transmission path refers to a combination of resources composed of perception, communication, computing, and storage resources in a specific order.} between the virtual source node and the virtual computing node. However, due to the conflict relationship in (14)-(16), these transmission paths cannot be directly called to transmit data. Fig. 5 (b) is the resource coupling relationship graph after the link scheduling algorithm. It's not difficult to understand that there are no resource conflicts in the graph, so we can schedule the data transmission directly on this graph.

\begin{figure}
\begin{center}
\includegraphics[width=3.2in,height=2.6in]{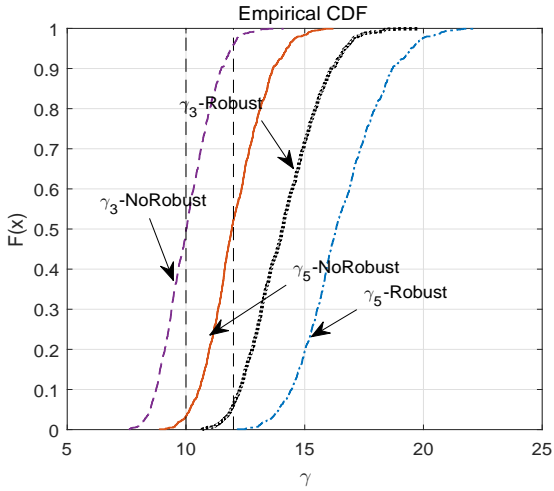}
\label{fig2}\caption{AVs SINR cumulative distributions.}
\end{center}
\end{figure}

Fig. 6 shows the learning results of the uncertain channel set. As it can be seen from the figure, the constructed ellipsoid set can surround the distribution of uncertain channels with a very high probability. If the obtained power allocation solutions are feasible under the constructed uncertainty set, the solution can satisfy the reliability constraint with a very high probability. Fig. 7 illustrates the AVs SINR cumulative distributions for different robust optimization approaches under the test set. In this experiment, the subchannels of AV3 and AV5 are reused by the V2V links $(1,2)$ and $(4,5)$, respectively. Moreover, the SINR requirements of the two AVs  are set as 10 and 12, respectively. It is not difficult to understand that the points $(10, F(10))$ and $(12, F(12))$ can be considered as the outage probabilities of AVs communications. Then, we observe from the figure that no matter which vehicle, the outage probability of Norobust approach is almost greater than 0.5. Such terrible performance is very dangerous for the emergency communication applications in IoV. On the contrary, by substituting the chance constraint as the proposed robust counterpart, the robust optimization approach achieves satisfactory outage performance.

\begin{figure}
\begin{center}
\includegraphics[width=3.2in,height=2.6in]{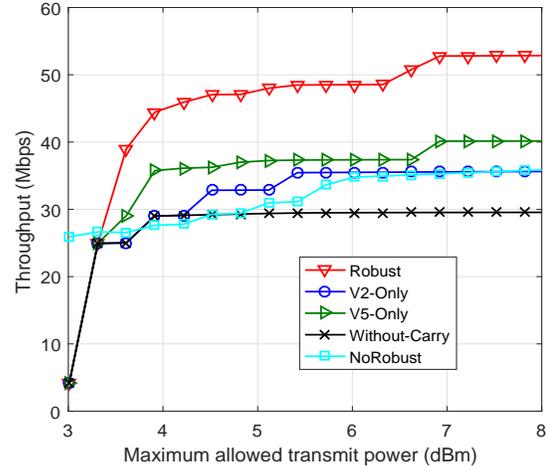}
\label{fig2}\caption{System throughput versus maximum allowed transmit power.}
\end{center}
\end{figure}

\begin{figure}
\begin{center}
\includegraphics[width=3.2in,height=2.6in]{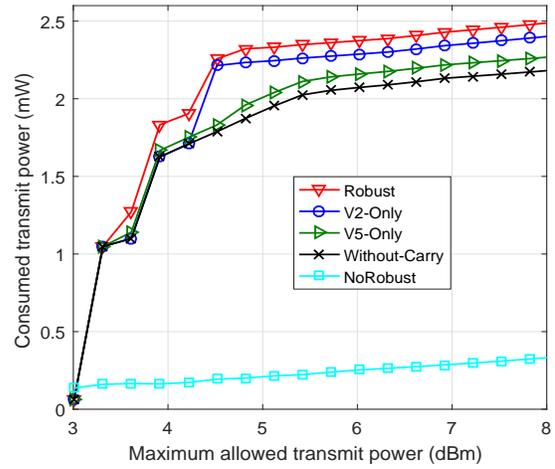}
\label{fig2}\caption{Consumed transmit power versus maximum allowed transmit power.}
\end{center}
\end{figure}

Figs. 8 and 9 show the system throughput and the consumed transmit power versus the system maximum allowed transmit power. We observe that with the increase of the maximum allowed transmit power, the throughput that the system can achieve increase. This is because the greater the transmit power, the greater the capacity of the communication arc. Therefore, the throughput of the system increases. However, limited by the capacity of the carry arc and computing arc, the throughput of the system increases to a certain extent and then stops increasing. It can also be seen from the figure that the system throughput achieved by the \emph{Without-Carry} scheme is the lowest. Even with the capacity of a single vehicle, the system throughput can be greatly improved. For example, both \emph{V2-Only}  and \emph{V5-Only} approaches achieve high throughput. This  demonstrates the benefits brought by the introduction of carry arc. Since our proposed robust scheme utilizes the carry capacity of v2 and v5 vehicles at the same time, it achieves the highest system throughput. However, in the \emph{NoRobust} scheme, the probability of network interruption is about 0.5, as shown in Fig. 7. Thus, the throughput it achieves is encumbered by the communication arc. This phenomenon verifies the significance of the robust power control in our proposed Robust scheme. As it can be seen in Fig. 9, in order to achieve the maximum system throughput, the proposed approach consumes the largest transmit power. Since the \emph{NoRobust} scheme only considers the average channel gain in the power allocation, thus it consumes the least transmit power. But the result is that it leads to the smallest throughput. It can also be seen from the figure that the consumed transmit power actually increases rapidly at first and then slowly at last. At the beginning, both the transmit powers of vehicles and BS are increasing with the increase of maximum allowed transmit power. When the maximum allowed transmit power reaches about 4.5, the size of traffic flow will be subject to the capacity of carry arc and computing arc, so it will not increase any more. Accordingly, the transmit power of the BS will not increase. However, the vehicle will continue to consume transmit power to maximize the capacity of the communication arc.  Therefore, the transmit power of the system will increase slowly.

\begin{figure}
\begin{center}
\includegraphics[width=3.2in,height=2.6in]{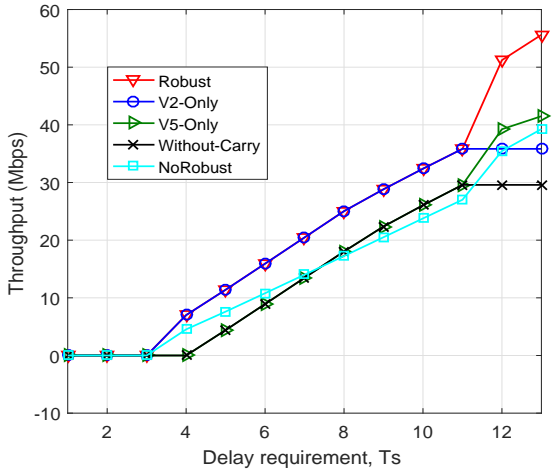}
\label{fig2}\caption{System throughput versus delay requirement.}
\end{center}
\end{figure}

\begin{figure}
\begin{center}
\includegraphics[width=3.2in,height=2.6in]{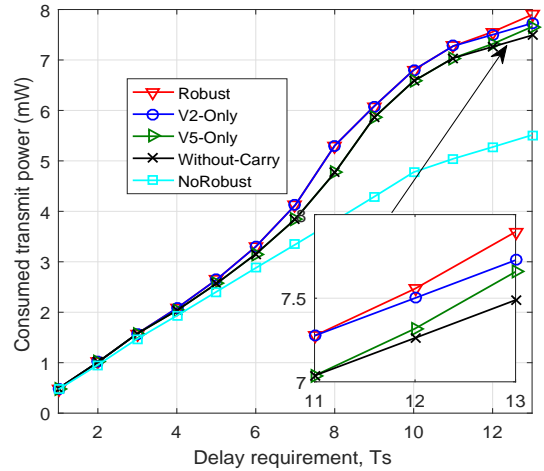}
\label{fig2}\caption{Consumed transmit power versus delay requirement.}
\end{center}
\end{figure}

Figs. 10 and 11 show the system throughput and the consumed transmit power versus the delay requirement. It can be seen from Fig. 10 that when the delay requirement is less than 3, the system throughput is always zero. This phenomenon can be explained together with Fig. 5(b). When the delay requirement is less than 3, there is no feasible path between the virtual source node and the virtual destination node. So the system throughput is always zero. When the delay requirement is 4, the cache of v2 can be used to construct several complete transmission paths. At the same time, from Fig. 10, we observe that \emph{Robust}, \emph{V2-Only} and \emph{NonRobust} approaches can achieve the throughput improvements. This again demonstrates the benefits of the introduction of carry arcs for the throughput improvement.
Because of the benefit of early use of the carry arcs, \emph{Robust} and \emph{V2-Only} approaches achieve higher system throughput when the delay requirement is greater than 3 and less than 11.
When the delay requirement is greater than 11, relay vehicle v2 can no longer connect with fog vehicle v3. Conversely, relay vehicle v5 can now be connected to fog vehicle v3. As a result, the throughput under \emph{v2-only} and \emph{Without-Carry} approaches no longer increases, while throughput under other approaches increases sharply. Fig. 11 shows the corresponding power consumption. As it can be seen from the figure, with the increase of delay requirement, more and more communication arcs can be utilized, so the system needs to allocate more transmit power to activate these links. When the delay requirement is greater than 11, although  \emph{V2-Only} and \emph{Without-Carry} approaches consume transmit power to activate link $(v_{5},v_{s})$, it does not carry data. Therefore, the BS does not consume power to carry the corresponding computing results. Conversely, in \emph{Robust} and \emph{V5-Only} approaches, activated link $(v_{5},v_{s})$ is capable of carrying data transmission, so the BS must consume power to transmit the corresponding computing results. Therefore, \emph{Robust} and \emph{V5-Only} approaches consume more transmit power when the delay requirement is greater than 11.

\begin{figure}
\begin{center}
\includegraphics[width=3.2in,height=2.6in]{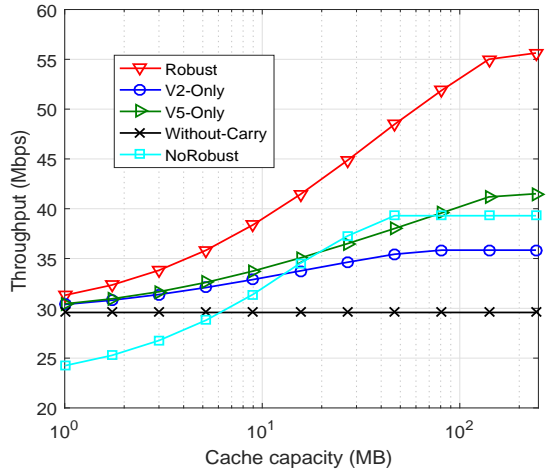}
\label{fig2}\caption{System throughput versus cache capacity.}
\end{center}
\end{figure}

\begin{figure}
\begin{center}
\includegraphics[width=3.2in,height=2.6in]{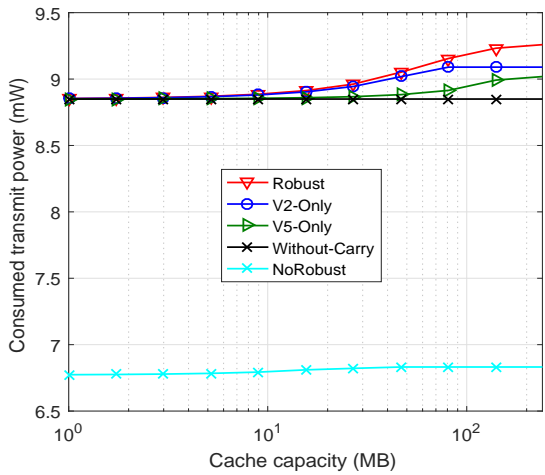}
\label{fig2}\caption{Consumed transmit power versus cache capacity.}
\end{center}
\end{figure}

Figs. 12 and 13 illustrate the system throughput and the consumed transmit power versus the cache capacity. As it can be seen from Fig. 12 that the system throughput under \emph{Robust}, \emph{V2-Only}, \emph{V5-Only} and \emph{NoRobust} approaches increase with the increase of cache capacity. The throughput achieved by \emph{Without-Carry} approach remains unchanged because it cannot utilize the carry arcs. We also find that our proposed scheme achieves maximum system throughput because it can take advantage of both v3 and v5 caches. By contrast, \emph{V2-Only} and \emph{V5-Only} approaches can only take advantage of one vehicle's cache, so they get less throughput than the \emph{Robust} one. Although \emph{NoRobust} approach can also take advantage of the cache of two vehicles, its communication link is unreliable, so it gains less throughput than the \emph{Robust} one and sometimes even less than \emph{Without-Carry}. Fig. 13 shows that the transmit power consumed by the system increases with the cache capacity. The reasons can be explained as follows. With the introduction of caching, the system can carry more data, so more power is required to support the data transmission.

\begin{figure}
\begin{center}
\includegraphics[width=3.2in,height=2.6in]{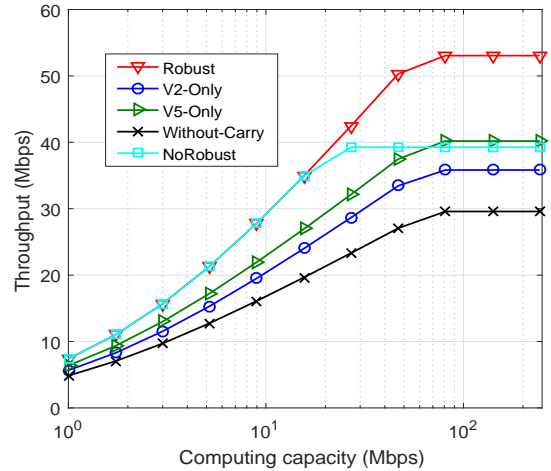}
\label{fig2}\caption{System throughput versus computing capacity.}
\end{center}
\end{figure}

\begin{figure}
\begin{center}
\includegraphics[width=3.2in,height=2.6in]{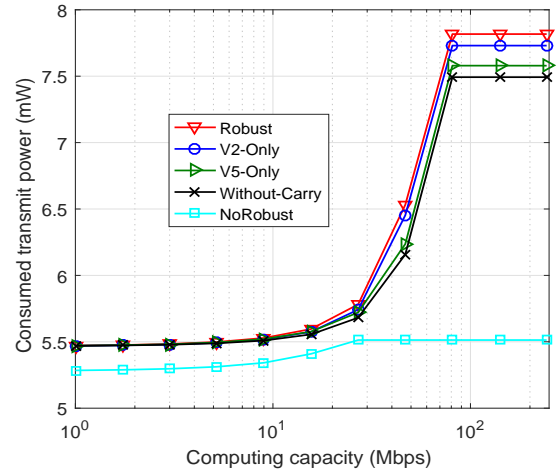}
\label{fig2}\caption{Consumed transmit power versus computing capacity.}
\end{center}
\end{figure}

Figs. 14 and 15 illustrate the system throughput and the consumed transmit power versus the computing capacity. It shows that the system throughput increases with the increase of computing capacity. This is because the more computing capacity a fog vehicle has, the more data the system can process. However, when the computing capacity reaches to 100Mbps, the throughput of the system does not increase any more. Clearly, the capacity of carry or communication arc becomes the bottleneck to throughput improvements. Fig. 15 shows the similar results. As the computing capacity increases, the system can process more data, which also requires more power to support the data transmission. When the data size stops increasing, there is no need for the system to consume power to support the data.

\section{Conclusions}

In this paper, we studied the cooperative content dissemination framework for IoV system. The intertwined impact of the objective perception, transmission, carry and computing was characterized by the time-expanded graph. Based on this graph model, we formulated the content dissemination process as a mathematical optimization problem, with the consideration of flow equilibrium constraint and the wireless capacity constraint. It should be noted that all possible communication paradigms, i.e., connected forwarding, carry-and-forward and direct forwarding, that can be implemented in vehicular networks were incorporated in the formulation of the optimization problem. Finally, we developed a cascaded joint link and subchannel scheduling algorithm and as well a robust joint power allocation and flow control algorithm for the content dissemination in IoV systems.

\footnotesize
\bibliographystyle{IEEEtran}
\bibliography{references}

\end{document}